\documentclass{aastex}
\usepackage{emulateapj5}

\begin{document}

\submitted{ApJL in press.}

\title{An Origin of Supersonic Motions in Interstellar Clouds}

\author{Hiroshi Koyama}
\affil{Astronomical Data Analysis Center, \\
National Astronomical Observatory, Mitaka, Tokyo 181-8588, Japan; \\
E-mail address: H.Koyama@nao.ac.jp }

\and

\author{Shu-ichiro Inutsuka}
\affil{Department of Physics, Kyoto University, Kyoto
606-8502, Japan; \\
E-mail address: inutsuka@tap.scphys.kyoto-u.ac.jp }

\begin{abstract}
The propagation of a shock wave into an interstellar medium 
is investigated 
by two-dimensional numerical hydrodynamic calculation with cooling, 
heating and
 thermal conduction. 
We present results of the high-resolution two-dimensional
 calculations to follow the fragmentation due to the thermal
 instability in a shock-compressed  layer. 
We find that geometrically thin cooling layer behind the shock front
 fragments into small cloudlets. 
The cloudlets have supersonic velocity dispersion in the warm neutral medium 
in which the fragments are embedded as cold condensations. 
The fragments tend to coalesce and become larger clouds.
\end{abstract}

\keywords{ISM: clouds
--- ISM: molecules
--- ISM: supernova remnants
--- shock waves
--- turbulence}

\section{INTRODUCTION}
The interstellar medium (ISM) and cold gas clouds are characterized by a
clumpy substructure and a turbulent velocity field (Larson 1981). 
The maintenance and dissipation processes of the turbulence are supposed
to be important in the theory of star formation (McKee 1989, Nakano
1998, Elmegreen 1999).
The understanding of the origin of cold clouds and their internal
substructure has therefore fundamental importance for a consistent theory
of star formation and ISM.

Some of the polarization maps and direct
measurements of field strength in some star forming regions suggested
the importance of MHD (Alfv\'{e}n waves) turbulence (Myers \&
Goodman 1988, Crutcher et al. 1993).
Recent simulations of MHD turbulence, however, suggest that it
dissipates rapidly  (Gammie \& Ostriker 1996, Mac Low et al. 1998, 
Mac Low 1999, Ostriker et al. 1999).
Possible sources of energy supply are winds and outflows from young
stellar objects (Franco \& Cox 1983, McKee 1989).
Note however that clumpy structures with supersonic
velocity dispersions are also observed
even in regions where star formation is inactive.  
Thus, the origin of clumpy cloud
structure remains as an outstanding issue. 

We propose that the clumpiness in clouds arises naturally from their
formation through thermal instability which acts on timescales that
can be much shorter than the duration of the interstellar shocks (e.g.,
galactic spiral shocks and supernova remnants).
The basic properties of the thermal instability are studied in a
pioneering paper by Field (1965). Schwarz, McCray, and Stein 
 (1972) studied numerically the growth of condensation in cooling region
including the effects of ionization and recombination.
Hennebelle \& Perault (1999) studied the elementary condensation process 
in turbulent flow in the restricted conditions of neutral atomic
gas in plane-parallel geometry. Burkert \& Lin (2000)
studied cooling and fragmentation of gas using simplified power-law
cooling function. Smith (1980) studied collisions between cold atomic clouds, 
which produce thick layers of shock-heated atomic gas in which thin
sheets of cold molecular gas form by the thermal instability. Koyama \&
Inutsuka (2000, hereafter Paper I) have done one-dimensional
hydrodynamic calculations for the propagation of a strong shock wave
into warm neutral medium (WNM) and cold neutral medium (CNM) including
detailed thermal and chemical processes. They have shown that the post-shock
region collapses into a cold layer as a result of the thermal instability.
They expect that this layer will break up into very small cloudlets which
have different translational velocities.

In this paper, we show that the fragmentation of the shock-compressed layer
indeed provides turbulent condensations,
by using two-dimensional hydrodynamic calculation with radiative cooling and
heating and thermal conduction. 

\section{NUMERICAL SIMULATIONS}
\subsection{Numerical Scheme}
The hydrodynamics module of our scheme is based on the second-order
Godunov method (van Leer 1979).
We solve following hydrodynamic equations.
\begin{equation}
\frac{\partial \rho}{\partial t}+\nabla\cdot(\rho {\bf v})=0,
\end{equation}
\begin{equation}
\frac{\partial {\bf v}}{\partial t}+{\bf v}\cdot\nabla{\bf v}=
-\frac{\nabla P}{\rho}, 
\end{equation}
\begin{equation}
\frac{\partial\rho e}{\partial t}
+\nabla\cdot(\rho e {\bf v})+P\nabla\cdot {\bf v}=
\frac{\rho}{m_{\rm H}}\Gamma-\left(\frac{\rho}{m_{\rm H}}\right)^2\Lambda(T)
+\nabla\cdot(K\nabla T),
\end{equation}
where $\rho$, $P$, ${\bf v}$ are the density, pressure, and velocity of
the gas, the specific internal energy $e=P/(\gamma-1)\rho$, with
$\gamma=5/3$.
$m_{\rm H}$ is the hydrogen mass in gram.
$\Gamma$ and $\Lambda$ are heating and cooling rate,
respectively. 
For the coefficient of thermal conductivity we adopt $K=2.5\times
10^3T^{1/2} {\rm erg \ cm^{-1}K^{-1}s^{-1}}$ (Parker 1953).  
Effects of self-gravity and
magnetic field are not treated in this paper.

In Paper I, we included the following processes in the hydrodynamic
calculations: 
photoelectric heating from small grains and PAHs, 
heating and ionization by cosmic rays and X-rays,
heating by H$_2$ formation and destruction,
atomic line cooling from hydrogen Ly$\alpha$, 
\ion{C}{2}, \ion{O}{1}, \ion{Fe}{2}, and \ion{Si}{2},
rovibrational line cooling from H$_2$ and CO, 
and atomic and molecular collisions with grains.
However the following analytic fitting function adopted here reproduces
the features relevant for our purpose:   

\begin{eqnarray}
\Lambda(T)/\Gamma&=&10^7\exp\left(-114800/(T+1000)\right) \nonumber \\
&&+14{\sqrt{T}}\exp\left(-92/T\right) {\rm cm^{3}}, \\
\Gamma&=&2\times 10^{-26} {\rm ergs/s},
\end{eqnarray}
where $T$ is in Kelvins.

We use 2048 $\times$ 512 Cartesian grid points covering a 1.44 $\times$
0.36 pc region so that the spatial resolution is 0.0007 pc = 140 AU.

We have tested the linear growth of the thermal instability (Field
1965). The test is performed with an eigenfunction by using 16 grid
points. The numerical results reproduce the linear growth rate with
relative errors of 0.05 \% when $\delta \rho/\rho <0.01$

\subsection{Initial and Boundary Conditions}

To investigate the shock propagation into WNM we consider a plane-parallel
shocked layer of gas.
We take the $y$-axis to be perpendicular to the layer.
An uniform flow approaches the layer from the upper side with a velocity 
$V_{\rm y}=-26$ km/s  and density $n=0.6 \ {\rm cm^{-3}}$ and temperature
$T=6000$ K.
We assume that the lower side of the
shocked layer is occupied by hot, tenuous gas (density $n=0.12 \ {\rm
cm^{-3}}$) with a high pressure 
$P/k_{\rm B}=4\times 10^4  {\rm K cm^{-3}}$.

We set up density fluctuations as follows:
\begin{equation}
\frac{\delta\rho(x,y)}{\rho_0}=
\frac{A}{k_{\rm max}}\sum_{{\rm i},{\rm j}=0}^{{\rm i}_{\rm max}}
\left({k_{\rm i}^2+k_{\rm j}^2}\right)^{\frac{n}{2}}
\sin(x k_{\rm i}+y k_{\rm j} + \theta(k_{\rm i},k_{\rm j})),
\end{equation}
where $k_{\rm i}=2\pi i/L$ is a wave number,
$\theta$ is a random phase, $n$ is a spectral index, and $L$ is a
calculation box size of $x$- direction.
We imposed out-going boundary condition for the $y$- direction, and 
periodic boundary condition for the $x$- direction.

\subsection{Results of Thermal Instability and Fragmentation}

\begin{figure}
\figurenum{1}
\epsscale{0.3}
\plotone{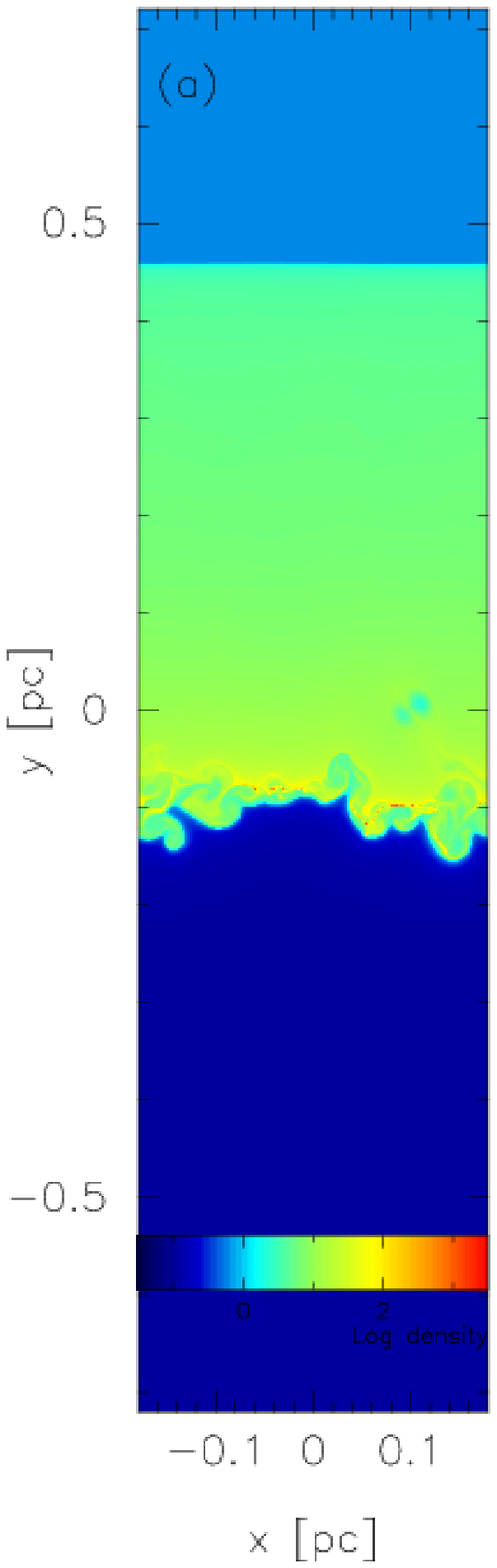}
\plotone{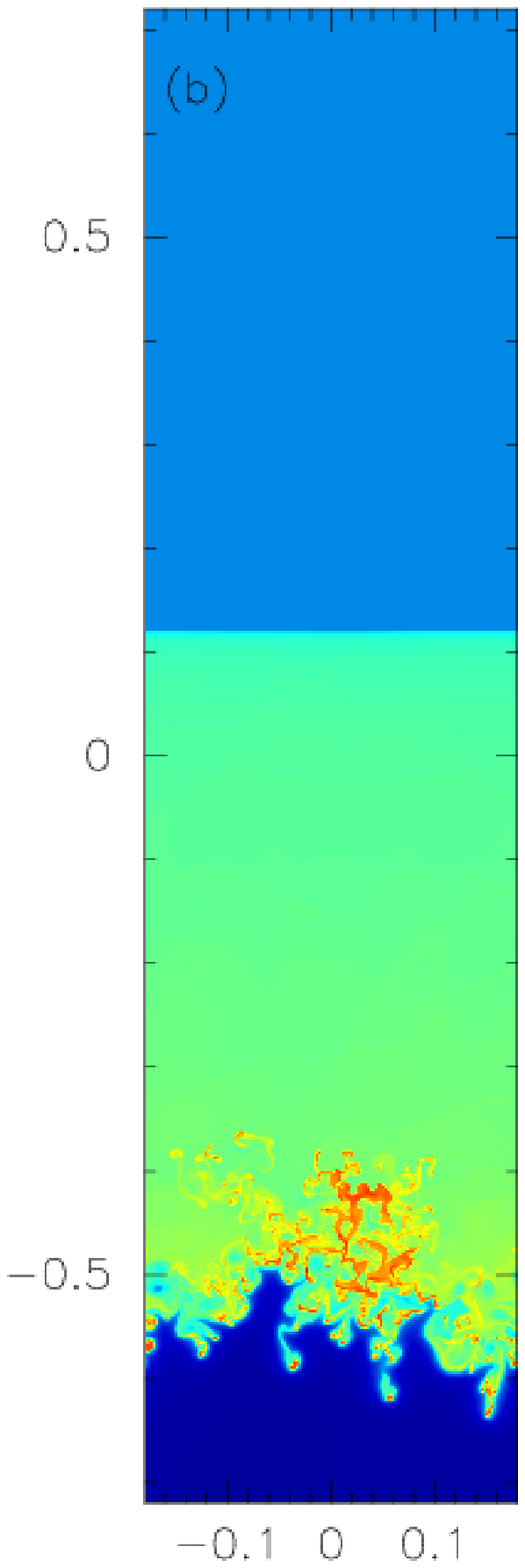}

\caption{Density distribution of the shock-compressed layers.
The figures show the snapshot at (a) $t=0.337$ Myr, (b) $t=1.06$ Myr,
respectively.}
\label{fig1}
\end{figure}

Figure \ref{fig1} shows density distribution at $t=$0.337, 1.06 Myr. 
We used an initial fluctuation $A=0.05$, the power index $n=0$, and
${\rm i}_{\rm max}=16$.
Thermally collapsed layer and its fragments are seen in the
shock-compressed layers.

Figure \ref{fig1}a shows the beginning of the
formation of cold thermally collapsed layer. 
Density discontinuity at $y\sim 0.46$ pc corresponds to the shock front.
Density discontinuity at $y\sim -0.1$ pc is the contact surface between
the shock-compressed layer and the hot interior.  
Behind the shock front, cooling dominates heating 
 and temperature decreases monotonically. 
The ISM in the range of 300 -- 6000 K is thermally unstable.
Thus, thermally collapsing layer is subject to thermal instability. 
The condensations continue to collapse and cool until the radiative cooling
balances the radiative heating, in such a way that approximate pressure
equilibrium is 
maintained between condensations and the surrounding gas.
Temperature of these clumps finally reaches the value of thermal
equilibrium ($T=20$ K at $n=2000 {\rm cm^{-3}}$).
When the cooling balances the heating, the condensation ceases to
collapse but the 
surrounding warm gas continues to cool radiatively and accrete onto the
cloud cores.

The cold clumps have considerable translational velocity dispersion.
The typical velocity dispersion is about several km/s.
The linear analysis predicts that perturbation with sufficiently long
wavelength grows exponentially.
However, the velocity of the non-linearly developed perturbation
has the upper limit that is the sound speed of the warmer medium
($\approx$ 10km/s),
because the driving force of the instability is the pressure of the
less dense warmer medium.
This non-linearly developed perturbation produces the translational
velocities of the cold clumps.
These velocity on the order of the sound speed of the WNM is
highly supersonic with respect to the sound speed of cold gas.
Thus we can understand why the supersonic velocity dispersion of
the cold medium is comparable to the sound speed of the WNM.

The cold cloud cools the interface via thermal conduction.
The pressure of the warm medium between two
cold clumps becomes smaller, which in effect provides an attractive
force of two adjacent cold clumps. This attractive force is the driving
force of the coalescence of two cold clumps.
Thus, cold clumps coalesce into larger clouds.

The sizes of these clumps are a few orders of magnitude smaller than the Jeans
length, $\lambda_{\rm J}=1.2 \ {\rm pc}\sqrt{(T/20{\rm K})(2000{\rm
cm^{-3}}/n) }$ . 
Thus, each cloud is gravitationally stable.
The shock front itself remains stable at least in $10^6$ years in this
plane-parallel problem.

For comparison, we changed the initial fluctuations in the range of the
amplitude (A=0.05--0.005), and the power index (n=0,$\pm$ 2), but the
resultant velocity dispersion defined by $\int \rho v_{rm x}^2\,dxdy/\int 
\rho\,dxdy$ changed only by 10 \%. We also changed the spatial
resolution (512x128, 1024x256, and 2048x512) but the velocity dispersion 
changed only by 10 \%. Thus, Figure \ref{fig1} shows a typical result of 
this problem.
From these results we can demonstrate that the role of thermal instability
behind the shock front is one of the most important processes 
in the evolution of interstellar shocked layers.

\section{DISCUSSION}

\subsection{Formation of Molecules}

The main coolant in cold dense clouds is supposed to be CO (see, e.g.,
Neufeld, Lepp, \& Melnick 1995; Paper I).
H$_2$ enables the formation of CO and other molecules.
The abundance of H$_2$ in the CNM critically depends on the optical
depth and local density.
The typical column density of the layer becomes 
$2\times 10^{20} {\rm cm^{-2}}$. 
We calculated detailed thermal-chemical equilibrium state in the dense
region by adopting the values of density, temperature, and column
density from this dynamical simulation.
Twenty-five percent of the
hydrogen is H$_2$ and 0.03\% of the
carbon is CO in thermal and chemical equilibrium.

\begin{figure}
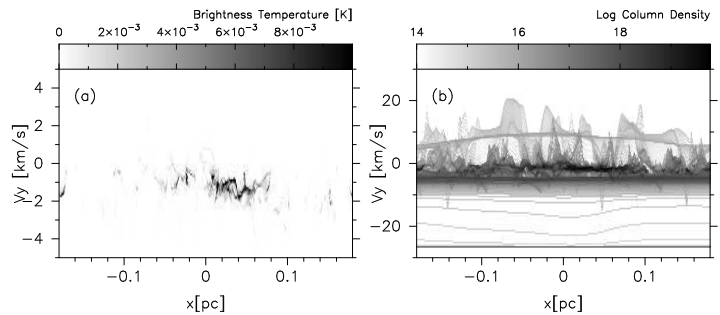

\figurenum{2}
\epsscale{0.25}
\plotone{fig2a.ps}
\plotone{fig2b.ps}

\caption{The panel a shows the position-velocity (P-V) diagram
 of the ${}^{12}$CO J=1 -- 0 emission obtained from the calculation at
 $t=1.06$ Myr (see Figure \ref{fig1}b ). We
 assume 0.03 \% of the carbon is CO. The panel b shows the P-V
 diagram of the hydrogen nuclei.}
\label{fig2}
\end{figure}

We simulate the observation of molecular clumps at $t=1.06$ Myr 
(Figure \ref{fig1}b) as the Position-Velocity (P-V)
diagram of the ${}^{12}$CO J = 1 -- 0 emission (Figure \ref{fig2}a).  
We use the usual relationship $T_{\rm B}=\frac{1}{2}\lambda^2I/k_{\rm
B}$, with the specific intensity $I=n^2\Lambda L/(4\pi \Delta v)$, where
$n^2\Lambda$ is the cooling per unit volume, $L$ is the path length,
$\Delta v$ is the Doppler line width. We adopt $\Delta v$ to be the
sound velocity. 
As shown in Figure \ref{fig2}a, the typical velocity dispersion in the
clumps is about a few km/s. The cloud to cloud velocity dispersion is
also a few km/s.
Figure \ref{fig2}b shows that the P-V diagram of hydrogen nuclei.
The velocity dispersion of the diffuse warm medium is about 10 km/s.

\subsection{An Origin of ``Turbulence'' in Interstellar Clouds}
Diffuse ISM in the Galaxy is frequently compressed by supernova
explosions (McKee \& Ostriker 1977).
Therefore, the shock propagation into the ISM plays an important
role in the evolution of the Galactic ISM. 
We have studied the shock propagation into WNM by two-dimensional
hydrodynamic calculations, and have shown that the thermally collapsed layer
breaks up into very small cloudlets.
Fragmentation of the thermally collapsing layer is a result
of the thermal instability. 
The thermal instability produces many cloudlets which have different
translational velocities. 
We expect that the Galactic ISM is occupied by these small
cloudlets which have supersonic velocity dispersion
, because the ISM is frequently compressed by supernova
explosions, stellar winds, spiral density waves, cloud-cloud
collisions, etc.
The \ion{H}{1} 21-cm observation maps show the existence of many
shell-like or filamentary structures in the Galaxy (Hartmann \& Burton
1997). In addition, high resolution observations suggested that the clumpy
distribution is ubiquitous in the Galaxy (Heiles 1997).
If these structures are really the results of the shock waves, many
small molecular cloudlets should be hidden in the shock-compressed
layers. These cloudlets should have translational velocities owing to the
fragmentation of thermal instability.
Observational ``turbulence'' in the ISM should reflect these motions.
Thus, we propose that an origin of ``turbulence'' in the ISM is the
motion of this small cloudlet complex. 

In the radiative shocked layers, the gases lose thermal energy through
radiative cooling.
Thus, the initial kinetic energy of the pre-shock gas (in the comoving
frame of the post-shock gas) converted to the radiation energy, which
escapes from the system.
If, however, the post-shock gas becomes dynamically unstable
by the thermal instability as in this paper,
some portion of the thermal energy is transformed into
the kinetic energy of the translational motions of the cold cloudlets,
which does not easily escape from the system.
Therefore we can consider the origin of interstellar turbulence is
due to the conversion of the gas energy in supersonic motion.
In principle, the kinetic energy of the cloudlets can be lost
via coalescence of the cloudlets.
The damping rate of the velocity dispersion of the cloudlets
will play a key role in the evolution of this system,
and hence, should be studied in the subsequent paper.

Expected effects of magnetic field modify the development of the thermal
instability described here.
The presence of magnetic field can suppress thermal conduction
efficiently, allowing the collapse of small scale structure by the thermal
instability.  
The magnetic forces suppress the motion that is perpendicular to the
magnetic field lines, and hence suppress the growth of perturbations
whose wavevector is perpendicular to the magnetic field lines.
However, 
the magnetic forces do not affect the motion that is parallel to the magnetic
field lines, and hence the growth of perturbations whose wavevector is
parallel to the magnetic field lines. 
As a result, the dense sheets or filaments will form and
tend to align {\em perpendicular} to the magnetic field lines.
We need
three-dimensional calculation to analyze these effects of magnetic fields.

Numerical computations were carried out on VPP300/16R and VPP5000/48 at the
Astronomical Data Analysis Center of the National Astronomical Observatory,
Japan.

\end{document}